\documentclass{article}
\usepackage[utf8]{inputenc}

\usepackage{geometry}
\usepackage{amsmath}
\usepackage{amssymb}
\usepackage{natbib}
\usepackage{color}
\usepackage{url}
\usepackage{graphicx}
\usepackage{booktabs}
\usepackage{rotating}
\usepackage[linesnumbered,ruled,vlined]{algorithm2e}
\usepackage{setspace}

\providecommand{\keywords}[1]
{
  \small	
  \textbf{\textit{Keywords---}} #1
}

\title{Bayesian scalar-on-tensor regression using the Tucker decomposition for sparse spatial modeling finds promising results analyzing neuroimaging data}
\author{Daniel Spencer, Rajarshi Guhaniyogi, Russell T. Shinohara, Raquel Prado}
\date{\today}

\begin{document}
	
	\maketitle
	
	\doublespacing
	
	\begin{abstract}
	    Modeling with multidimensional arrays, or tensors, often presents a problem due to high dimensionality. In addition, these structures typically exhibit inherent sparsity, requiring the use of regularization methods to properly characterize an association between a tensor covariate and a scalar response. We propose a Bayesian method to efficiently model a scalar response with a tensor covariate using the Tucker tensor decomposition in order to retain the spatial relationship within a tensor coefficient, while reducing the number of parameters varying within the model and applying regularization methods. Simulated data are analyzed to compare the model to recently proposed methods. A neuroimaging analysis using data from the Alzheimer's Data Neuroimaging Initiative shows improved inferential performance compared with other tensor regression methods.
	\end{abstract}
	
	
	\section{Introduction} \label{sec:Introduction_tucker}


Neuroimaging analysis is an important application area due to the broad societal impact of brain disorders, which account for significant decline in quality of life, in addition to billions of dollars in cost of care and lost work for a large portion of the population \citep{olesen2003burden,diluca2014cost,feigin2021burden}. Modeling neuroimaging data presents a particular challenge due to high dimensionality and low signal-to-noise present in current brain imaging modalities. Often, the goal in the analysis of these data is to make population-level inferences about the function of the brain and the pathology of specific brain disorders using high-resolution structural brain images taken from living subjects. Due to the clinical importance of inference from these models, assumptions and constraints must be carefully applied. 

Several methods are already in use for these types of data within the neuroimaging community. One of the most commonly-used methods is referred to as the general linear model (GLM), which is not to be confused with the generalized linear model that is commonly used in statistics. The GLM performs a massive univariate analysis in which the response is regressed independently on each voxel within a tensor covariate, in addition to any additional vector covariates \citep{friston1995spatial,penny2011statistical}. This model has the advantage of being relatively computationally inexpensive. However, the GLM also assumes that the associations between different voxels in the tensor and the response are independent and no assumptions are made about the underlying sparsity of the tensor coefficient, which negatively impacts model power. In practice, different multiple testing corrections are used to preserve spatial relationships among neighboring voxels, though work by \cite{eklund2016cluster} suggests that these inflate the false discovery rate. Methods that control the false discovery rate are appealing for their interpretability \citep{benjamini1995controlling, lindquist2015zen}, but they fail to take spatial relationships within the tensor coefficient into account.

Another class of models that have been applied to imaging data are spatially-varying models such as Gaussian Markov Random Field (GMRF) models. \cite{goldsmith2014smooth} proposed a model using an Ising prior with a GMRF prior to enforce sparsity and smoothness in the estimates of the spatially-varying estimates of the nonzero coefficients. However, this model relies on cross-validation to set parameters that control smoothness on the parameter estimates in the prior, and the use of binary indicators leads to mixing concerns in the MCMC. \cite{chiang2017hierarchical} added spatially-informed indicators of nonzero parameters in a classification model that also included a study of functional connectivity for inference at a region-of-interest level, but this would still be expected to suffer from mixing difficulties in the MCMC chain. In addition, these models suffer from a very large parameter space that would be computationally challenging to implement at the scale of the models within the neuroimaging literature.

Envelope models, such as those proposed by \cite{cook2010envelope}, reviewed by \cite{lee2020review}, and extended by \cite{zhang2017tensor}, among others, achieve drastic dimension reduction, but require high signal-to-noise ratios or relatively large sample sizes in order to achieve satisfactory estimation. 

A different class of approaches takes advantage of the tensor structure of the data by decomposing the tensor covariates using one of two tensor decompositions and imposing regularization constraints on the tensor factors. The advantage of these models is that they allow for computationally feasible inference at the voxel-level. Work done by \cite{zhou2013tensor} uses the parallel factorization/canonical polyadic (PARAFAC/CP) tensor decomposition in a classical tensor regression approach, which assumes that dimension margins are independent components for construction of the tensor coefficients. This was expanded in work by \cite{li2018tucker} in the use of the more flexible Tucker decomposition, which allows for interaction effects between different decomposition rank elements. However, identifiability constraints in the model can induce false positive detection. \cite{guhaniyogi2017bayesian} created a novel Bayesian prior structure on the PARAFAC/CP tensor decomposition elements, which improved on the uncertainty quantification in the model inference. Investigation into Bayesian priors with the CP decomposition was extended to include inference on correlation using a Gaussian graphical model in \cite{spencer2020joint} and further explored in \cite{guhaniyogi2021bayesian} with a multiway stick-breaking shrinkage prior, but these are done in the context of tensor-on-scalar regression on time series data.

We propose a Bayesian scalar-on-tensor regression model using the Tucker decomposition (BTRT) that assumes sparsity and spatial dependence within a tensor-valued coefficient in order to improve estimation over previous scalar-on-tensor models while using Bayesian methods to provide uncertainty quantification. This is accomplished through the use of the Tucker tensor decomposition \citep{tucker1966some}, which can be thought of as an analog to an independent components analysis in which the components are dimension margins. In addition, we accomplish effective and accurate modeling without imposing identifiability constraints other than the shrinkage priors on the tensor decomposition elements. We show the efficiency and accuracy of the model over other tensor decomposition methods through simulation studies. The BTRT model is then used to perform a neuroimaging analysis of data from the Alzheimer's Disease Neuroimaging Initiative\footnote{Data used in preparation of this article were obtained from the Alzheimer’s Disease Neuroimaging Initiative
(ADNI) database (adni.loni.usc.edu). As such, the investigators within the ADNI contributed to the design
and implementation of ADNI and/or provided data but did not participate in analysis or writing of this report.
A complete listing of ADNI investigators can be found at:
\url{http://adni.loni.usc.edu/wp-content/uploads/how_to_apply/ADNI_Acknowledgement_List.pdf}}. The BTRT model and tensor regression competitors are implemented in the R software package \texttt{bayestensorreg} \citep{spencer2022bayestensorreg}.

We formally define terminology and use of tensor notation, and explain the BTRT and competitor models in section \ref{sec:Methodology}. Inferential accuracy is examined through the use of simulated datasets in section \ref{sec:simulateddata}. An analysis of data from the Alzheimer's Disease Neuroimaging Initiative in section \ref{sec:AnalysisADNI} illustrates improvements to inferential power using real data. We end with a discussion of our findings and directions for future research in section \ref{sec:Discussion}.

\section{Methodology} \label{sec:Methodology}

This section begins with the introduction of tensor notation and the formal definition of two tensor decompositions. Next, the BTRT model and its prior structure is described. Model identifiability is then discussed, followed by a guide on how to choose the model rank. Finally, competitor models are outlined, which validate our proposed method as a useful and reliable tool in sparse tensor regression scenarios.

\subsection{Tensor terminology and notation}

A tensor of order $D$ is a multi-dimensional array data structure $\mathbf{A} \in \mathbb{R}^{p_1,\ldots,p_D}$. To illustrate, a vector is a tensor of order 1, a matrix is a tensor of order 2, a cubic array is a tensor of order 3, and so forth. 

The \textit{vectorization} of a tensor $\mathbf{A} \in \mathbb{R}^{p_1 \times \cdots \times p_D}$ of order $\geq 2$ results in a tensor of order 1 of length $\prod_{j=1}^D p_j$, that is $\text{vec}(\mathbf{B}) \in \mathbb{R}^{p_1 \ldots p_D}$. The \textit{inner product} of two tensors $\mathbf{A}$ and $\mathbf{A}'$ is the crossproduct of the vectorized elements of the tensors, that is $\langle \mathbf{A}, \mathbf{A'} \rangle  = \text{vec}(\mathbf{A})^\top\text{vec}(\mathbf{A'})$. 

The $k$th-mode \textit{matricization} of a tensor, represented as $\mathbf{A}_{(k)}$ is a matrix representation of a tensor of order 2 or higher such that the $k$th index becomes the first index and all other tensor indices are combined in order into a second index. That is, $\mathbf{A}_{(k)} \in \mathbb{R}^{p_k \times p_1 p_2 \cdots p_{k-1} p_{k+1} \cdots p_{D}}$.

Tensors can be approximated via different \textit{tensor decompositions}. One such decomposition is called the canonical decomposition/parallel factorization, also known as CANDECOMP/PARAFAC, or CP \citep{tucker1966some}. This decomposition represents the tensor $\mathbf{B}$ as 
\begin{align}
\mathbf{B} = \sum_{r=1}^R \boldsymbol{\beta}_{1,r} \circ \cdots \circ \boldsymbol{\beta}_{D,r}, \label{eq:cp_decomposition}
\end{align}
in which $\boldsymbol{\beta}_{j,r} \in \mathbb{R}^{p_j}$ is one of $R$ principal components for the $j$th dimension of $\mathbf{B}$. Here, $R$ is known as the \textit{rank} of the CP decomposition. The $\circ$ operator denotes the \textit{outer product}. At each value of $r \in \{1,\ldots,R\}$, the series of outer products $\boldsymbol{\beta}_{1,r} \circ \cdots \circ \boldsymbol{\beta}_{D,r}$ results in a $D$-dimensional tensor summand $\mathbf{B}_r \in \mathbb{R}^{p_1,\ldots,p_D}$. As the value of $R$ increases, the number of summands in the decomposition increases and the resolution improves. All tensors can be represented exactly by the CP decomposition given sufficiently high rank $R$. In practice, low-rank decomposition approximations have been found to be adequate to estimate sparse tensor coefficients with spatial dependence. Using the CP decomposition reduces the parameter space for estimating $\mathbf{B}$ from $\prod_{j = 1}^D p_j$ to $R\sum_{j=1}^D p_j$, while incorporating this sparse spatial structure. However, it is possible that not all dimension margins in a tensor decomposition require all $R$ components in order to faithfully represent structure within a tensor. In such cases, the CP decomposition may be extended to the \textit{Tucker decomposition}, which can be written as 
\begin{align}
\mathbf{B} = \sum_{r_1 = 1}^{R_1} \cdots \sum_{r_D = 1}^{R_D} g_{r_1,\ldots,r_D} \boldsymbol{\beta}_{1,r_1} \circ \cdots \circ \boldsymbol{\beta}_{D,r_D}, \label{eq:tucker_decomposition}
\end{align}
where
$$ \mathbf{G} = (g_{r_1,\ldots,r_D})_{r_1=1,\ldots,r_D = 1}^{R_1,\ldots,R_D} \in \mathbb{R}^{R_1 \times \cdots \times R_D}$$
is the \textit{core tensor} composed of elements, which assigns weights of importance to each of the $\prod_{j = 1}^D R_j$ tensor summands that compose the tensor $\mathbf{B}$ \citep{tucker1966some}. This representation is more flexible than the CP decomposition, as it allows for different decomposition ranks for different tensor dimension margins, and all ranks of different margin factors interact with every rank in other dimension margins. As an example, consider an order 2 tensor. A rank 2 CP decomposition uses two tensor summands to approximate the tensor, while a rank 2, 2 Tucker decomposition uses a weighted sum of four tensor summands at the cost of only four additional model parameters from the core tensor. The parameter space for estimating $\mathbf{B}$ using the Tucker decomposition is $\prod_{j = 1}^D R_j + \sum_{j=1}^D R_jp_j$.

With these basic conventions, a linear model can be built using the Tucker decomposition that is more flexible with only a modest increase in the number of parameters in the parameter space.

\subsection{Tensor Regression Model}
Assuming a scalar response $y_i$, vector covariates $\boldsymbol{\eta}_i \in \mathbb{R}^q$, and tensor covariates $\mathbf{X}_i \in \mathbb{R}^{p_1 \times \cdots \times p_D}$ for subjects $i = 1,\ldots,n$, the observed linear model can be represented as 
\begin{align}
y_i & = \langle \mathbf{B},\mathbf{X}_i \rangle + \boldsymbol{\gamma}' \boldsymbol{\eta}_i + \epsilon_i, \label{eq:linear_model}
\end{align}
in which $\mathbf{B} \in \mathbb{R}^{p_1 \times \cdots \times p_D}$ is a tensor coefficient, $\boldsymbol{\gamma} \in \mathbb{R}^q$ is a vector coefficient, and $\epsilon_i$ is an error term, which follows any distribution centered at zero. Note that, since we are focusing on the methods to address tensor regression, we are restricting our attention to scenarios in which $q$ is relatively small. However, the model framework here does allow for a seamless extension to high-dimensional modeling of $\boldsymbol{\eta}_i$, e.g. for genetic modeling.

In the potentially large predictor space within the tensor-valued covariate, it is reasonable in certain applications to assume spatial dependence in the association between $\mathbf{X}_i$ and $y_i$. In order to  impose this structure in the model while simultaneously reducing the parameter space, we use the Tucker tensor decomposition, as outlined in (\ref{eq:tucker_decomposition}). We use Bayesian methods to define assumptions about the sparsity of the tensor coefficient using state-of-the-art regularization priors. As the other terms in the model are not the focus of our proposal, typical Bayesian modeling techniques are applied to the $\boldsymbol{\gamma}$ and $\epsilon_i$ terms, as outlined below.

\subsection{Bayesian Tensor Regression Tucker prior structure}

We focus our model on applications in which a degree of sparsity can be reasonably assumed in the tensor-valued coefficient. Classical regularization methods used to meet such a goal include various penalized regression algorithms like the LASSO \citep{tibshirani1996regression} or RIDGE regression \citep{hoerl1970ridge}. However, these methods lack the ability to provide a measure of uncertainty quantification on the estimates for the parameters, which is important for determining the practical significance of the estimates for such parameters. Shifting to a Bayesian modeling structure can serve to fill this gap, leading to improved inference. 

\cite{li2018tucker} proposes a classical model for tensor regression using the Tucker decomposition in which a penalty is applied to either the core tensor $\mathbf{G}$ or both the core tensor $\mathbf{G}$ and each dimension component $\boldsymbol{\beta}_{j,r_j}$ in the decomposition. We propose shrinkage priors on both the core tensor and all dimension components in order to more strongly induce parameter regularization. 

Following the previous work done by \cite{guhaniyogi2017bayesian}, an adapted generalized double-Pareto prior is applied to the dimension components within the Tucker tensor decomposition. That is,
\begin{align*}
\boldsymbol{\beta}_{j,r_j} & \sim \text{Normal}\left(  \mathbf{0}, \tau \mathbf{W}_{j,r_j} \right), &
\tau & \sim \text{Gamma}(a_\tau, b_\tau), \\
\omega_{j,r_j,\ell} & \sim \text{Exponential}\left( \frac{\lambda_{j,r_j}^2}{2} \right), &
\lambda_{j,r_j} & \sim \text{Gamma}(a_\lambda,b_\lambda),
\end{align*}
where $\mathbf{W}_{j,r_j}$ is a diagonal matrix with elements $\omega_{j,r_j,\ell}$ for $\ell = 1,\ldots,p_j$. Integrating over the element-specific scale parameters reduces the prior on $\beta_{j,r_j,\ell}$ to a double-exponential (Laplace) distribution centered at 0 with a scale parameter of $\frac{\lambda_{j,r_j}}{\sqrt{\tau}}$, which has heavier tails than a Gaussian distribution while also allocating higher densities around zero. This prior is a Bayesian analog to the adaptive LASSO, inheriting its desirable oracle properties of the maximum a posteriori estimator \citep{armagan2013generalized}. 

In order to adequately select the proper rank for each dimension and to reduce noise in the tensor coefficient estimates, the generalized double-Pareto prior is also imposed on the elements of the core tensor $\mathbf{G}$:
\begin{align*}
g_{r_1,\ldots,r_D} & \sim \text{Normal}(0,z v_{r_1,\ldots,r_D}), &
z & \sim \text{Gamma}(a_z,b_z), \\
v_{r_1,\ldots,r_D} & \sim \text{Exponential}\left( \frac{\varphi_{r_1,\ldots,r_D}^2}{2} \right), &
\varphi_{r_1,\ldots,r_D} & \sim \text{Gamma}(a_\varphi,b_\varphi).
\end{align*}
This combination ensures that only tensor summands within the Tucker decomposition that explain additional variance are included in the estimate of $\mathbf{B}$. 

For the purposes of our analyses, a multivariate normal prior with mean $\boldsymbol{\mu}_\gamma$ and covariance $\boldsymbol{\Sigma}_\gamma$ is placed on the elements of $\boldsymbol{\gamma}$. This is done to maintain conjugacy while maintaining control over the expected effects of the vector-valued coefficients. The errors  ($\epsilon_i$s) in this model are assumed to be independent identically distributed  following a normal distribution with a mean of zero, and a variance of $\sigma_y^2$. An inverse gamma prior is placed on $\sigma_y^2$ with hyperparameters $a_\sigma$ and $b_\sigma$ in order to maintain conjugacy.
\begin{align*}
\boldsymbol{\gamma} & \sim \text{Normal}(\boldsymbol{\mu}_\gamma, \boldsymbol{\Sigma}_\gamma), &
\epsilon_i & \overset{i.i.d.}{\sim} \text{Normal}(0,\sigma_y^2), &
\sigma_y^2 & \sim \text{Inverse Gamma}(a_\sigma, b_\sigma)
\end{align*}

The inference surrounding the vector-values parameters $\boldsymbol\gamma$ is not a central focus of our proposed model. However, $\boldsymbol\mu$ and/or $\boldsymbol\Sigma$ could have structures or additional hierarchical priors placed on them, depending on the nature of the data application.

\subsection{Identifiability}

The decomposition of the tensor coefficient affects the identifiability of the parameters in the model. That is, the value of any voxel within the tensor coefficient is estimated as 
$$\hat{b}_v \in \hat{\mathbf{B}} : \hat{b}_v = g_{1,\ldots,1} \beta_{1,1,v_1} \cdots \beta_{D,1,v_D} + \ldots + g_{R_1,\ldots,R_D} \beta_{1,R_1,v_1} \cdots \beta_{D,R_D,v_D}, $$
where $v$ is the voxel location within the tensor ($v = (v_1,\ldots, v_D)$). Any two of these summands can be multiplied by $c$ and $\frac{1}{c}$, respectively, and the estimate for that voxel within the tensor coefficient $\hat{b}_v$ would remain unchanged. Indeed, the values of $g_{r_1,\ldots,r_D}$ and $\beta_{j,r_j,v_j}$ are not identifiable, and the only identifiability constraint placed on these parameters is the regularization effect of the priors to influence the model to place higher posterior densities closer to zero. In contrast, the FTR models rely on setting the first $R$ or $R_j$ values equal to 1 for the CP and Tucker decompositions, respectively, which can result in larger coefficient estimates close to the image border.

In testing the BTRT model in very low signal-to-noise settings with scalar coefficients with values close to zero, models in which at least one margin rank $R_j = 1$ can result in divergent MCMC chains. This happens because core tensor values multiply with only the rank 1 tensor decomposition factor, which can have a multiplicative inverse effect on the value of the scalar coefficient. In our experiments, increasing all $R_j$ to be $\geq 2$ alleviated this identifiability issue.

Something to consider when using the model in \textbf{Equation (\ref{eq:linear_model})} for a data analysis is that an identifiability problem exists between $\mathbf{B}$ and $\boldsymbol{\gamma}$ if there are voxels within $\mathbf{X}_i, \, \forall i = 1,\ldots,n$ that are collinear with any of the values in $\boldsymbol{\eta}_i, \, \forall i = 1,\ldots,n$. This is a problem that may be undetected in two-step frequentist models with regularization, as either elements in $\mathbf{B}$ or $\boldsymbol{\gamma}$ may simply be assigned values of zero.

\subsection{Selection of Rank} \label{sec:RankSelection}

A key consideration in the use of the BTRT model structure is the selection of each dimension rank. Increases in a margin's rank can be made with an attempt to increase the spatial resolution on the inference of the tensor coefficient. For larger tensor covariates, this requires higher ranks if the nonzero coefficients are not arranged hypercubically. Choice of unequal ranks may be prudent if some of the tensor dimensions are much larger or smaller than others. Consider an example in which the tensor covariate has dimensions $100 \times 100 \times 4$. The ranks for the first two dimensions may need to be considerably larger than the rank for the third dimension, as there are only four possible margin locations in the final dimension. One such realistic application would be to combine magnetic resonance images that use different sequences (e.g. T1 weighted, T2 weighted, effective T2, etc.) into a single tensor, using a low rank on the dimension that represents different sequences. This is a clear advantage over the CP/PARAFAC decomposition methods, as each dimension is not forced to have the same number of ranks as all of the others. 

Importantly, the ranks for different tensor margins are not exchangeable. That is, if one has fit a BTRT model with ranks (1, 2); then this will not necessarily yield the same results as a BTRT model with ranks (2, 1).

Finding the optimal rank for the model fit starts by fitting  BTRT models with $R_j = r;\, \forall j$ for increasing values of $r$ until the deviance information criterion (DIC) \citep{gelman1995bayesian} for the model stops decreasing. Refer to the value of $r$ that minimizes the DIC for this set of models as $R^*$. Next, fit the models with ranks such that one margin has rank $R^* - 1$ and all other margins maintain rank $R^*$. If one of these models produces a lower DIC than the model in which $R_j = R^*;\, \forall j$, set that model as the new baseline with ranks $(R_1^*,\ldots,R_D^*)$. For each dimension margin $j$, fit the model with all of the the baseline margin ranks except with $R_j = R_j^* - 1$. The model that produces a smallest DIC value from these models then becomes the new baseline rank model. Continue this process to find the combination of ranks that minimizes DIC. For example, the rank search path in a two dimensional model could be 
$$(1,1) \longrightarrow (2,2) \longrightarrow (3,3) \longrightarrow (4,4) \longrightarrow (3,2);(2,3) \longrightarrow (3,1) \longrightarrow (3,2),$$
resulting in selecting the model with ranks (3, 2) after fitting models for 7 different rank combinations.


\subsection{Competitor Models}

The effectiveness of the proposed Bayesian sparse tensor regression models is shown by making direct comparisons to models commonly used in the field of neuroscience. The first is the general linear model (GLM), in which the response $(y_1,\ldots,y_n) = \mathbf{y}$ is regressed on each voxel $v$ within the tensor covariate $\mathbf{X} \in \mathbb{R}^{p_1 \times \ldots \times p_D \times n}$ independently of the vector-valued coefficients. The linear model for the GLM is written in equation (\ref{eq:GLM}). Note that the same responses $(y_1,\ldots,y_n) = \mathbf{y}$ are used to fit separate models for each voxel within the tensor covariate.
\begin{align}
y_i & = \boldsymbol{\gamma}'\boldsymbol{\eta}_i + \epsilon_{i,v}, \nonumber  \longrightarrow
\tilde{y}_i = y_i - \hat{\boldsymbol{\gamma}}'\boldsymbol{\eta}_i, \nonumber \\
\tilde{y}_i & = b_vX_{i,v} + \epsilon_{i,v}' \label{eq:GLM}
\end{align}
This model is an industry standard for its ease of implementation and high computational efficiency. However, this model often suffers from a high false discovery rate in multiple-subject analyses \citep{eklund2016cluster}. We adjust for the high false discovery rate in our implementation by using the Benjamini-Hochberg multiple testing correction \citep{benjamini1995controlling}, which fixes the false discovery rate (FDR). In the following trials, the FDR is fixed at 0.05, in agreement with standard practice in neuroimaging.

Direct tensor regression competitors are also used to validate the performance of the proposed model structure. Specifically, the frequentist tensor regression using the PARAFAC/CP tensor decomposition (FTR CP) \citep{zhou2013tensor} and the Tucker tensor decomposition (FTRT) \citep{li2018tucker} are used to compare point estimates to classical methods. These methods apply the LASSO to the margin factors in their respective tensor decompositions in \textbf{Equations (\ref{eq:cp_decomposition}) and (\ref{eq:tucker_decomposition})}. The FTRT model also allows for the application of the LASSO to the elements of the core tensor. In our tests, the FTR CP model performs best with the LASSO applied to the margin factors and the FTRT model performs best when the LASSO is only applied to the elements within the core tensor, so these are the configurations used in our comparisons. 

We also compare to the Bayesian tensor regression using the PARAFAC/CP tensor decomposition (BTR CP) \citep{guhaniyogi2017bayesian} provides a more direct comparison in terms of point estimates and uncertainty quantification. 

\section{Simulated Data Analysis} \label{sec:simulateddata}

In order to demonstrate the efficacy of the BTRT model, data were simulated from the linear model in (\ref{eq:linear_model}) under the following conditions. $\mathbf{B} \in \mathbb{R}^{50 \times 50}$, where nonzero-valued elements take the value of 1 in the middle of the regions, decreasing to lower nonzero values using the \texttt{specifyregion} function within the \texttt{neuRosim} package in R \citep{neuRosim}. For the sake of these simulations, three separate regions of activation are generated at random locations. The elements of $\mathbf{X}_i$ are all independently generated from a standard normal distribution for $i = 1,\ldots, 1000$. The elements for the vector-valued covariates $\boldsymbol{\eta_i}$ are also independently generated from a standard normal distribution. The parameters for the vector-valued covariates are set as $\boldsymbol{\gamma} = (\gamma_1, \gamma_2, \gamma_3) = (25, 3, 0.1)$ to show how the different models estimate parameters of different size, relative to the observation error, which is set to have a variance of 1. Finally, the elements of $\mathbf{y} = (y_1,\ldots,y_n)$ are generated according to \textbf{Equation (\ref{eq:linear_model})} where $\epsilon_i \overset{iid}{\sim} \text{Normal}(0,1)$. 

The Bayesian models had the following hyperparameter settings: $a_\sigma = 3$, $b_\sigma = 20$, $a_\lambda = 3$, $b_\lambda = a_\lambda^{1/(2D)}$, $\boldsymbol{\mu}_\gamma = \mathbf{0}$, $\boldsymbol{\Sigma}_\gamma = 900\mathbf{I}_q$, $a_\tau = 1$, $b_\tau = \min(R_1,\ldots,R_D)^{(1/D) - 1}$, $a_z = 1$, $b_z = \min(R_1,\ldots,R_D)^{(1/D) - 1}$, $a_\varphi = 3$, and $b_\varphi = a_\varphi^{1/(2D)}$, where the order of the coefficient tensor is set to $D=2$. The prior on the observation error variance is set to be relatively noninformative in the context of the simulation, with a mean of 10 and a variance of 100. The priors for $\lambda$ and $\varphi$ are set to have modes between 1.5 and 2, approaching 2 as the tensor dimension increases, which places the prior expected value for $v_{r_1,\ldots,r_D}$ and the elements within $\mathbf{W}_{j,r_j}$ to be between $\frac{2}{3}$ and $\frac{1}{2}$. The priors for $\tau$ and $z$ have a mean of 1 when all $R_1,\ldots, R_D = 1$, increasing sublinearly with both the minimum value of $R_j$ for $j = 1,\ldots,D$, and the tensor dimension $D$. The prior variance for $\tau$ and $z$ increases linearly in rank when $D=2$, and superlinearly in rank when $D>2$. This prior specification allows for a slightly higher prior variance for $\boldsymbol{\beta}_{j,r_j}$ and $g_{r_1, \ldots, r_D}$ as rank and dimension increase in order to allow for moderately faster exploration of the parameter space as the tensor dimension and rank increase. The prior for $\boldsymbol{\gamma}$ is set to be relatively noninformative.


For all BTR models, 1,000 Markov Chain Monte Carlo (MCMC) simulations from the posterior distribution were drawn. After discarding the first 1000 draws, the remaining draws were used to estimate $\mathbf{B}$ and $\boldsymbol{\gamma}$. Following the rank search algorithm in section \ref{sec:RankSelection}, the optimal rank for the BTRT models is found to be the model with ranks (4,4).

Continuous distribution priors do not produce parameter estimates equal to zero. In order to perform variable selection, the sequential 2-means variable selection method proposed by \cite{li2017variable} is used to find point estimate values in $\mathbf{B}$. The method works by using 2-means clustering of any subset of the parameter space on the absolute values of each draw from the posterior distribution. The number of values in the cluster in which the center is furthest from zero is taken as the number of non-zero valued parameter estimates for that particular posterior sample. The median number of non-zero parameter estimates across all of the posterior samples, $m$ is then found. Finally, the parameters with the $m$ highest posterior median absolute values are determined to have true non-zero values. These parameters are then estimated with their posterior medians. This process is formally described in \textbf{Algorithm \ref{alg:sequential2means}}.

\begin{figure}[htp]
	\centering
	\includegraphics[width=\textwidth]{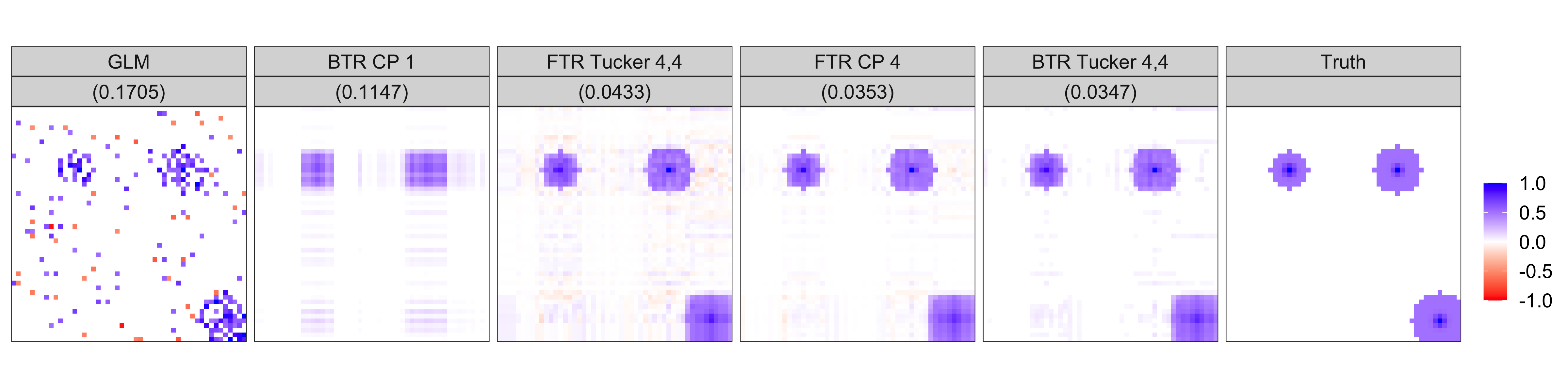}
	\caption{Point estimates of the true tensor coefficient for all competitors. The Bayesian models were chosen using the deviance information criterion, and the frequentist models were chosen using the Akaike information criterion. Values for the root mean squared error for each estimate are shown in parentheses, with lower values indicating better estimates.}
	\label{fig:sim_point_estimate_competitors}
\end{figure}

\begin{figure}[htp]
    \centering
    \includegraphics[width=\linewidth]{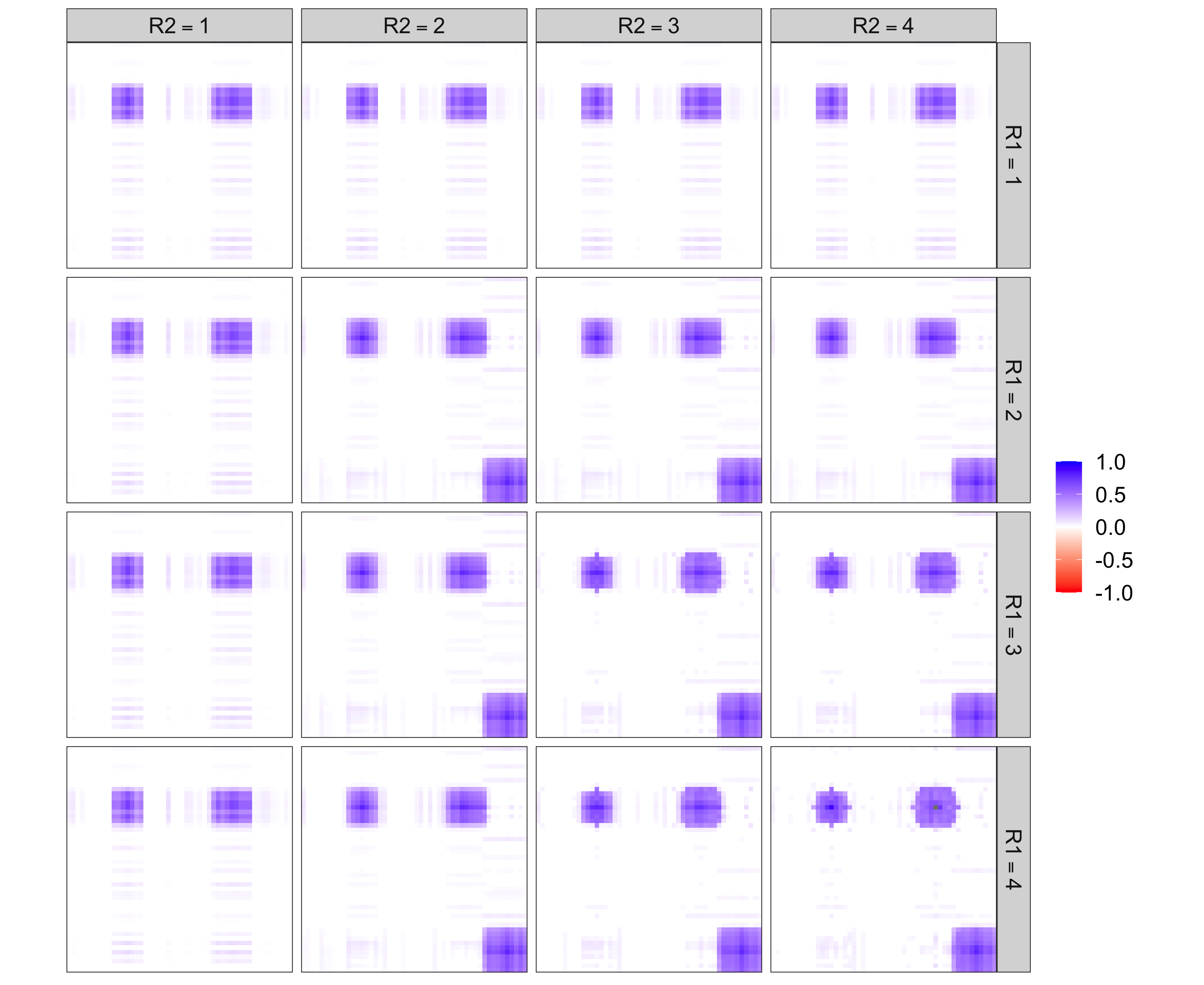}
    \caption{Point estimates for the image coefficient in the simulated data for the BTR Tucker model.}
    \label{fig:simulated_btrt_B}
\end{figure}

Point estimates for the tensor coefficient from the Bayesian tensor regression (BTR) models, frequentist tensor regression (FTR) models, and the GLM are shown in \textbf{Figure \ref{fig:sim_point_estimate_competitors}}. The ranks of the models shown were selected using the DIC for the Bayesian models or the Akaike's Information Criterion (AIC) for the frequentist models. We choose to use the AIC for the frequentist model selection because it is analogous to the DIC for the Bayesian models. For comparison, the Bayesian information criterion (BIC), which more heavily penalizes for overparameterization, selects the FTRT rank 2,2 model and the FTR CP rank 4 model. The estimate shown for the GLM tensor coefficient assigns values of zero to voxels within the tensor that are not statistically significant after using the Benjamini-Hochberg multiple testing correction with a Type I error rate of 0.05. Included in this figure are the values for the root mean squared error, defined as $\sqrt{\frac{1}{V} \sum_v (\hat{B}_v - B_v)^2}$, where $\hat{B}_v$ is the point estimate for tensor coefficient element $v$ obtained from a model, $B_v$ is the true value taken by the coefficient tensor element $v$, and $V$ is the total number of elements in the coefficient tensor. These results show that all models except for the BTR CP model choose higher model ranks in order to capture the shapes and magnitudes of the three distinct nonzero regions in the tensor coefficient. Higher ranks are necessary in order to capture multiple nonzero regions, but the BTR CP model shows increased posterior variance as model rank increases, which biases the rank selection for that model toward lower values. The BTRT estimate is both visually and numerically most similar to the true values in this simulated data analysis. Notably, all of the tensor regression models perform significantly better than the two-step GLM model, which is often used in neuroimaging studies. \textbf{Appendix \ref{app:rmse_allBTRT_sim}} contains \textbf{Table \ref{tab:sim_rmse_B}}, which gives the RMSE values for all ranks of all competing models, and \textbf{Figure \ref{fig:simulated_btrt_B}}, which shows the point estimates for all ranks of the BTRT model.

\begin{figure}[htp]
	\centering
	\includegraphics[width=\linewidth]{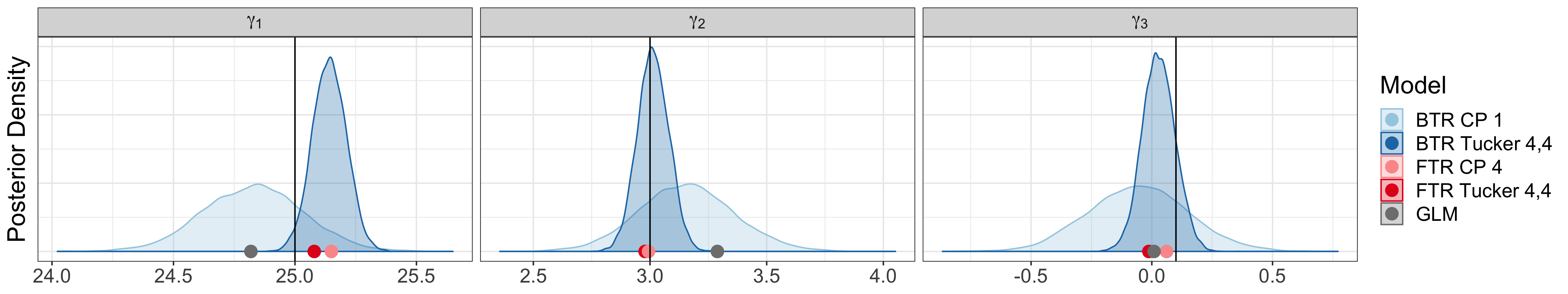}
	\caption{Posterior densities for \{$\gamma_1, \gamma_2, \gamma_3$\} from the selected Bayesian models. Points indicate the estimates from the selected frequentist sparse tensor regression models or the general linear model (GLM). The black line indicates the true value.}
	\label{fig:gamma_densities}
\end{figure}

The posterior densities for \{$\gamma_1, \gamma_2, \gamma_3$\} under different models with selected ranks can be seen in Figure \ref{fig:gamma_densities}. These results show that the posterior densities for each $\gamma_j$ are centered at or around the true values. The BTRT model shows much higher posterior densities around the true coefficient values than the BTR CP model, an effect of the improved estimation of $\mathbf{B}$. Though the BTR CP rank 1 model is chosen by the DIC, a comparison between the BTR CP rank 4 model and the BTRT rank 4,4 model still shows that the BTRT model is able to concentrate the posterior density for $\boldsymbol\gamma$ closer to the true values than the BTR CP model. In general, the posterior modes for the BTRT model are in agreement with the point estimates from the FTR models, suggesting that the model performs well in the estimation of the vector coefficients.


\subsection{Model Convergence}

In Bayesian modeling settings, it is important to ensure that the MCMC converges around an area within the neighborhood of the global posterior mode. Given that the posterior inference matches the inference shown within the frequentist models, the model does converge to a global mean in the simulated data settings when the assumption of normal error is satisfied, and the prior distributions are specified to be relatively uninformative. The log-likelihood values using the posterior draws, shown in \textbf{Figure \ref{fig:simLLIK}} in \textbf{Appendix \ref{app:rmse_allBTRT_sim}}, also show rapid convergence to a mode and posterior stability. The median effective sample sizes, as defined in \cite{gong2014practical}, across all voxels in $\mathbf{B}$ range from about 5400 to about 8100 out of 10000 draws from the posterior distribution, indicating sufficient numbers of effective samples from the posterior distribution were taken to make reliable inference.

\subsection{Hyperparameter Sensitivity}

Modeling in such a high-dimensional space with several hierarchical levels in the prior structure requires careful selection of hyperparameter values. However, the model is still expected to be somewhat robust to differences in prior specification. In order to test this expectation, a large grid of hyperparameter values for $a_\sigma, b_\sigma, a_\lambda, a_\tau, a_z$, and $a_\varphi$ was created. First, the values for the hyperparameters used to model the simulated data are taken and multiplied by 0.1, 1, or 10. One hundred configurations were randomly selected from the 729 possible configurations, and the model with ranks $R_1 = R_2 = 3$ was run with the same simulated data as above for 11,000 iterations with each combination. In each case, 1,000 iterations in the MCMC were discarded as a burn-in. 

Of the 100 configurations, 93 converged to have an RMSE for the tensor coefficient $\mathbf{B}$ between 0.03279 and 0.03336. The remaining 7 configurations all had RMSEs within $10^{-8}$ of 0.1336523, which is the RMSE that results from predicting $\mathbf{B}$ with a tensor of all zeros. The priors for these seven configurations all had the smallest values for the prior variance for the elements of $\mathbf{B}$ (approximately $2 \times 10^{-3}$) and/or the highest prior means (approximately $10^{10}$) and variances (over $10^{20}$) for $\sigma_y^2$. Indeed, individual inspection of these models with larger RMSE values show that the estimates for $\boldsymbol{\gamma}$ are within the same range as the estimates from the models that had a lower RMSE for the tensor coefficient, but they all estimate that the tensor coefficient is zero. This highlights the need to carefully balance the amount of shrinkage that can be applied to elements within $\mathbf{B}$ so that the values do not shrink all the way to zero, but they are also not strongly influenced by the prior to take nonzero values, if that would be inaccurate. In addition, the prior specification for $\sigma_y^2$ must be chosen so that nonzero values in the model coefficients for $\mathbf{B}$ and $\boldsymbol\gamma$ can be detected without allowing values to be inflated by placing strong prior weight on high observed variance. In general, these problems can be avoided by centering and scaling the response, and using the hyperparameter values given at the beginning of this section.

\section{Analysis of ADNI data} \label{sec:AnalysisADNI}

Structural differences in the brain can be measured using tensor-based morphometry (TBM), which provides a measure for the amount of deformation that needs to be applied to each voxel in order to match a given template image \citep{brun2010statistically}. Larger values indicate that the neighborhood around a given voxel is larger relative to the standard to which subject structural scans were registered, and smaller values indicate smaller relative volumes. These measures have been used to deduce volumetric phenotypes for many diseases, such as Alzheimer's disease and HIV/AIDS \citep{lepore2006multivariate}. 

Data used in the preparation of this article were obtained from the Alzheimer’s Disease Neuroimaging Initiative (ADNI) database (\url{adni.loni.usc.edu}). The ADNI was launched in 2003 as a public-private partnership, led by Principal Investigator Michael W. Weiner, MD. The primary goal of ADNI has been to test whether serial magnetic resonance imaging (MRI), positron emission tomography (PET), other biological markers, and clinical and neuropsychological assessment can be combined to measure the progression of mild cognitive impairment (MCI) and early Alzheimer’s disease (AD).

In our analysis, we apply the BTRT method and competing methods to measure associations between a subject's performance on the mini mental state exam (MMSE) and TBM data in the ``TBM Jacobian Maps MDT-SC" collection from the Alzheimer’s Disease Neuroimaging Initiative (ADNI) database \citep{hua2013unbiased}, which includes data for 817 subjects (342 female) from the first phase of the ADNI  (ADNI-1). All data were accessed April 21, 2022. Of these subjects, 188 were diagnosed with early Alzheimer's, 400 subjects exhibited mild cognitive impairment, and 229 were healthy controls. The TBM images from the dataset are relative to a structural template in $\mathbb{R}^{220\times 220 \times 220}$ created for the analysis of these data by \cite{hua2008tensor}. For illustration, we examine TBM data from the 80th axial slice, which includes subcortical structures such as the amygdala and the hippocampus, which previous studies have suggested are associated with Alzheimer's disease and cognitive decline \citep{barnes2009meta,poulin2011amygdala}. This slice image contains $220^2 = 48,400$ voxels. It is possible to analyze the whole brain using the current implementation of the software package, but future work to improve computational speed will greatly reduce model runtimes. Phenotype data were also included for each subject in the analysis, matched to patient ID  using the \texttt{ADNIMERGE} package in R \citep{adnimerge}.

The MMSE is a diagnostic tool used to classify adults based on levels of cognitive impairment \citep{pangman2000examination}. The exam itself poses a series of questions to test the subject's ability to perform everyday tasks, i.e. reading an analog clock or reproducing a drawn shape. The maximum score that one can achieve on the test is 30, and the scores take integer values. These scores can then be used in conjunction with other information to diagnose an individual. MMSE values for the subjects in this analysis ranged from 18 to 30.

We include years of education, and the number of Apolipoprotein E4 (APOE4) genetic variants as additional covariates for each subject following exploratory data analysis of the scalar variables available within the ADNI. APOE4 has been identified as a genetic risk factor for Alzheimer's disease \citep{strittmatter1996apolipoprotein}, and this variable can take the values 0, 1, or 2. We treat the APOE4 covariate as continuous, as the exploratory data analysis suggests a linear association with the MMSE.

We compare the performance of the different models in this analysis with the root mean squared predictive error (RMSPE) and the Pearson correlation between the predictions and the actual values in the dataset. The Pearson correlation is included because it was used as a performance metric in the DREAM challenge held by ADNI \citep{zhu2016compass}. All BTRT models are run for 11,000 iterations using the same hyperparameter values specified for the simulated data analyses in \textbf{Section \ref{sec:simulateddata}}, after which 1000 iterations are discarded as a burn-in. The BTR CP model was successfully tested using simulated data under several different scenarios. However, MCMC chains for the TBM data fail to maintain numerical stability despite starting from several different initial conditions, and the BTR CP model was unable to produce more than about 30 samples from the posterior distribution for any model rank between 1 and 4. Thus, the BTR CP results are excluded from comparison here. For the BTRT model, the point predictions for the response are calculated as the medians of the posterior predictive distributions, which are found for each subject as
\begin{align}
\hat{y}_{i}^{(s)} & = \langle \mathbf{B}^{(s)},\mathbf{X}_i \rangle + \boldsymbol{\gamma}^{(s)'} \boldsymbol{\eta}_i, \label{eq:ppd}
\end{align}
for each sample $s$ from the posterior distribution. The frequentist tensor regression models are run until the relative log-likelihood change between steps is less than 0.1\%, and they predict the response value for each subject using their point estimates for $\mathbf{B}$ and $\boldsymbol{\gamma}$. Model ranks for the BTRT, FTRT, and FTR CP models are chosen following the guidelines in \textbf{Section \ref{sec:RankSelection}}, resulting in the rank 3,3 BTRT model, the rank 1,1 FTRT model, and the rank 1 FTR CP model. The median effective sample size for the elements within $\mathbf{B}$ was 10,000.

\begin{figure}[htp]
	\centering
	\includegraphics[width=\linewidth]{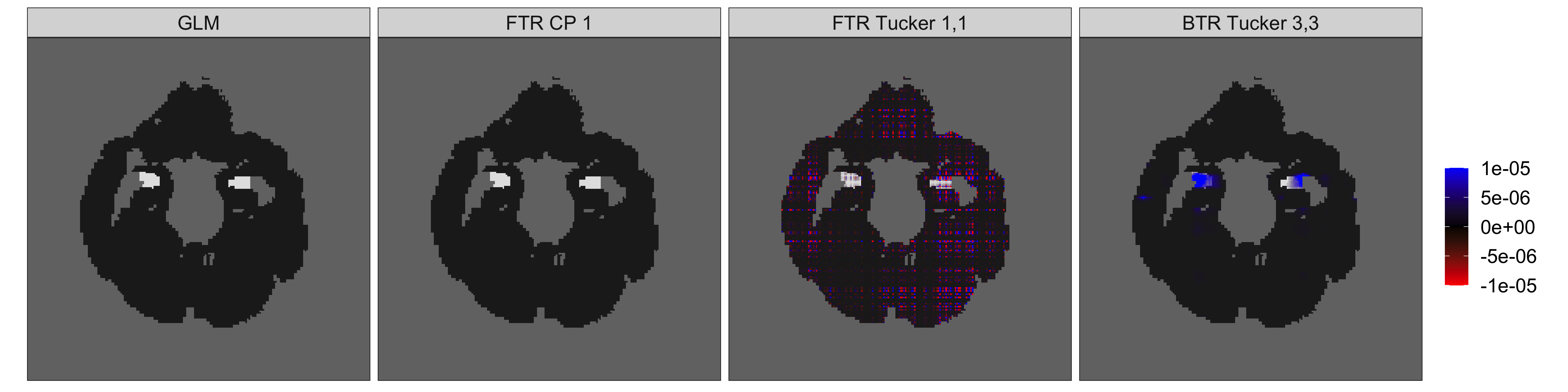}
	\caption{Point estimates of the image coefficient from the different models from the analysis of the TBM data from the Alzheimer's Disease Neuroimaging Initiative. The region highlighted in white corresponds to the amygdala within the automated anatomical labelling atlas.}
	\label{fig:tbm_best_B}
\end{figure}

Final point estimates for the tensor coefficient are found in the Bayesian models by using the sequential 2-means post-hoc variable selection method \citep{li2017variable}, which is outlined in \textbf{Appendix \ref{app:seq2means}}. The frequentist tensor regression models are not corrected for multiple testing, as they use the LASSO to select the coefficients that are significantly different from zero. The GLM model estimate is found by applying the Benjamini-Hochsberg multiple testing correction to the image coefficient p-values and setting voxels that are not significant to have values of zero. The DIC is used to select which BTRT model should be used to fit the data, and the AIC is used to select the rank of the FTR models, though the BIC selects the same rank FTR models in this analysis. The plots for the final estimates can be seen in \textbf{Figure \ref{fig:tbm_best_B}}. The estimates shown here have had their values thresholded to be between $\pm 1 \times 10^{-5}$ in order to show how models perform in terms of indicating areas of nonzero values, as they produce estimates that differ in orders of magnitude. We used the automated anatomical labelling atlas \citep{tzourio2002automated} to highlight the amygdala in the background image, as it has been implicated in previous studies of Alzheimer's disease \citep{poulin2011amygdala}. We mask the image coefficient estimates to be within the brain to match our inferential goal of finding associations within brain structures, rather than measures of deformation outside the brain that would correspond to associations with total brain volume.

Estimates for the image coefficient greatly vary among the different models, which may suggest a very low signal-to-noise ratio within these data to conclude that there are nonzero coefficient values. Nonetheless, the point estimate from the BTRT model clearly highlights positive associations between the relative size of the amygdala region and the MMSE scores. The GLM model finds no locations that are significantly different from zero after performing a multiple testing correction to limit the false discovery rate and using a Type I error rate of 0.05. The FTR CP model selected using the AIC also results in no nonzero values within the coefficient image. Thus, the GLM and the FTR CP models become equivalent to the model fitted without the image covariate in this analysis. The selected FTRT model does find nonzero values all over the brain, but the values found are both positive and negative and not spatially-specific. This suggests that the BTRT model furnishes superior inference over other tensor regression models in the low signal-to-noise ratio setting of real data.

\begin{figure}[htp]
	\centering
	\includegraphics[width=\linewidth]{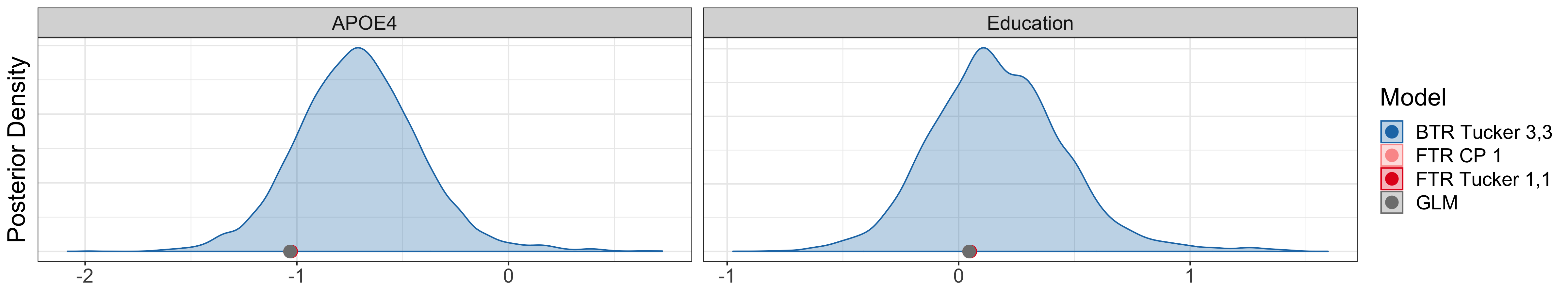}
	\caption{The posterior density for the values of the vector-valued covariates from the BTRT model, along with the estimates from the frequentist sparse tensor regression and general linear models.}
	\label{fig:tbm_gam}
\end{figure}

Posterior distributions for the non-tensor coefficients ($\boldsymbol\gamma$) are shown in \textbf{Figure \ref{fig:tbm_gam}}, along with the point estimates for the parameters from the GLM and the FTR models. These densities show an agreement between the competitor models and the BTRT model in the estimation of these coefficients in the BTRT model. In terms of model fitting, these estimates and densities could be used to make judgements about whether some covariates are valuable for making predictions or inference. The relatively small effect size and posterior variance for education variable suggest that it does not exhibit strong linear associations with MMSE scores given the associations of the TBM image and APOE4 coefficients.

The Pearson correlation and the RMSPE values are shown in \textbf{Table \ref{tab:RMSPE_gam_ADNI}}. We can see from this table that the BTRT model performs dramatically better in both measures while also providing interpretable estimates for the image coefficient.

%

\begin{table}
\centering
    \begin{tabular}{l|r|r}
    \hline
    \textbf{Model} & \textbf{Pearson} & \textbf{RMSPE}\\
    \hline
    No image / GLM & 0.273 & 2.574\\
    \hline
    FTR CP Rank 1 & 0.273 & 2.574\\
    \hline
    FTR Tucker Rank 1,1 & 0.276 & 2.571\\
    \hline
    BTR Tucker Rank 3,3 & \textbf{0.595} & \textbf{2.168}\\
    \hline
    \end{tabular}
    \caption{The root mean squared prediction error and Pearson correlation for fitted values. Note that the GLM produces no values that are significantly different from zero, so the model is equivalent to the model fitted without the image covariate.}
    \label{tab:RMSPE_gam_ADNI}
\end{table}

\section{Discussion} \label{sec:Discussion}

We propose a Bayesian tensor regression model using the Tucker tensor decomposition that accounts for spatial dependence through the structure of the prior without explicitly defining a Gaussian process or an approximation, which would have higher computational requirements. The proposed method and competitor models, as well as simulated data generation have all been implemented within the \texttt{bayestensorreg} R package \citep{spencer2022bayestensorreg}. We compare the performance of the Bayesian tensor regression model using the Tucker decomposition to other recent tensor regression models using simulated data to show improved inference and model fit. We also apply the BTRT model and competitors in the analysis of neuroimaging data from the ADNI project.

The BTRT model shows improvement over competing models within simulation studies. The BTRT models outperform the frequentist Tucker tensor regression models by more closely matching the true tensor coefficients visually and numerically via the root mean squared error metric. The selected BTRT model dramatically outperforms the BTR CP model in terms of image coefficient estimates due to lower variance within the MCMC chains, which remains consistent among different model ranks. The BTRT model also produces  posterior distributions with lower variance than the BTR CP model for the non-image-valued parameters centered around the true generated values, indicating improved inferential performance. All of the tensor regression models show a marked improvement over the general linear model, which is often used in neuroimaging literature.

Analysis of the TBM data from the ADNI illustrates the value of the BTRT model in providing meaningful inference while performing well on model reliability metrics. The results we found suggest that the model assumptions are able to detect sparse, spatially-dependent nonzero image coefficients better than the GLM and FTR models.

While the BTRT model performs impressively in terms of model fit on these two dimensional cases, additional work needs to be done in order to make this model a feasible option for fitting whole-brain volumes. Computational costs can be driven down using expectation-maximization or variational inference methods, and implementation in a more efficient programming language is also expected to improve efficiency.

\section{Acknowledgements}

Data collection and sharing for this project was funded by the Alzheimer's Disease Neuroimaging Initiative
(ADNI) (National Institutes of Health Grant U01 AG024904) and DOD ADNI (Department of Defense award
number W81XWH-12-2-0012). ADNI is funded by the National Institute on Aging, the National Institute of
Biomedical Imaging and Bioengineering, and through generous contributions from the following: AbbVie,
Alzheimer’s Association; Alzheimer’s Drug Discovery Foundation; Araclon Biotech; BioClinica, Inc.; Biogen;
Bristol-Myers Squibb Company; CereSpir, Inc.; Cogstate; Eisai Inc.; Elan Pharmaceuticals, Inc.; Eli Lilly and
Company; EuroImmun; F. Hoffmann-La Roche Ltd and its affiliated company Genentech, Inc.; Fujirebio; GE
Healthcare; IXICO Ltd.; Janssen Alzheimer Immunotherapy Research \& Development, LLC.; Johnson \&
Johnson Pharmaceutical Research \& Development LLC.; Lumosity; Lundbeck; Merck \& Co., Inc.; Meso
Scale Diagnostics, LLC.; NeuroRx Research; Neurotrack Technologies; Novartis Pharmaceuticals
Corporation; Pfizer Inc.; Piramal Imaging; Servier; Takeda Pharmaceutical Company; and Transition
Therapeutics. The Canadian Institutes of Health Research is providing funds to support ADNI clinical sites
in Canada. Private sector contributions are facilitated by the Foundation for the National Institutes of Health
(\url{www.fnih.org}). The grantee organization is the Northern California Institute for Research and Education,
and the study is coordinated by the Alzheimer’s Therapeutic Research Institute at the University of Southern
California. ADNI data are disseminated by the Laboratory for Neuro Imaging at the University of Southern
California.

\bibliography{Tucker.bib}
\bibliographystyle{plainnat}

\appendix

\newpage

\section{Other performance metrics for simulated data analysis} \label{app:rmse_allBTRT_sim}

\begin{table}[ht]
	\centering
	\begin{tabular}{l|r|r|r|r}
		\toprule
		& \multicolumn{4}{c}{FTR Tucker} \\
		\hline
		$R_1/R_2$ & 1 & 2 & 3 & 4\\
		\hline
		1 & 0.1166 & 0.1161 & 0.1161 & 0.1160\\
        \hline
        2 & 0.1167 & 0.0525 & 0.0524 & 0.0526\\
        \hline
        3 & 0.1167 & 0.0522 & 0.0478 & 0.0473\\
        \hline
        4 & 0.1167 & 0.0521 & 0.0464 & \textbf{0.0433}\\
		\hline
		& Rank 1 & Rank 2 & Rank 3 & Rank 4 \\
		\hline
		FTR CP & 0.1142 & 0.0492 & 0.0416 & \textbf{0.0353}\\
		\hline
		\cmidrule{1-5}
		& \multicolumn{4}{c}{BTR Tucker} \\
		\hline
		$R_1/R_2$ & 1 & 2 & 3 & 4\\
		\hline
		1 & 0.1143 & 0.1143 & 0.1144 & 0.1144\\
        \hline
        2 & 0.1143 & 0.0495 & 0.0495 & 0.0495\\
        \hline
        3 & 0.1144 & 0.0495 & 0.0411 & 0.0411\\
        \hline
        4 & 0.1144 & 0.0495 & 0.0411 & \textbf{0.0347}\\
		\hline
		& Rank 1 & Rank 2 & Rank 3 & Rank 4 \\ 
		\hline
		BTR CP & \textbf{0.1147} & 0.0494 & 0.0419 & 0.0364\\
		\hline
		& \underline{No Ranks} &  &  & \\
		GLM &  0.1705 &  &  & \\
	\end{tabular}
	\caption{The root mean squared error (RMSE) for the point estimates of the tensor coefficient from the simulated data. Values in bold font represent the model selected using the AIC for the FTR models and the DIC for the BTR models.}
	\label{tab:sim_rmse_B}
\end{table}

\begin{figure}[htp]
	\centering
	\includegraphics[width=\linewidth]{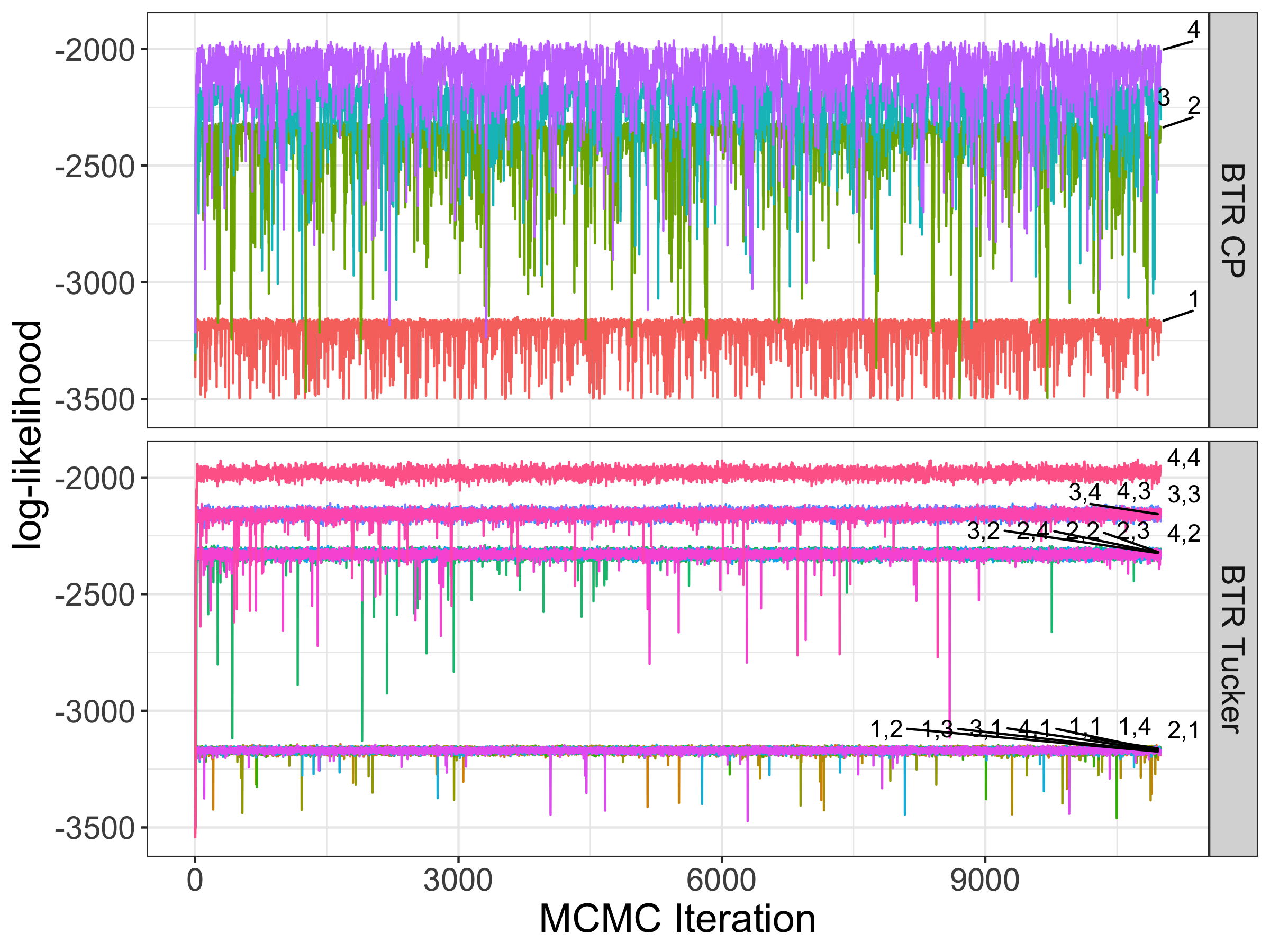}
	\caption{The log-likelihoods for the Bayesian models applied to the simulated data. Labels at the end of the lines give the model rank. These plots suggest rapid convergence of the MCMC in all models, though the variance in the log-likelihood from iteration-to-iteration is much lower in the Tucker decomposition models than in the CP/PARAFAC decomposition models.}
	\label{fig:simLLIK}
\end{figure}

\newpage

\section{Sequential 2-means algorithm}\label{app:seq2means}

\begin{algorithm}[ht]
	\SetAlgoLined
	\KwResult{Final estimate of $\boldsymbol{\theta}$ with small elements set to be equal to 0}
	\For{$s \gets 1$ \KwTo $S$}{
		Cluster the absolute value of elements in $\boldsymbol{\theta}^{(s)}$ into two clusters, $\mathcal{A}$ and $\mathcal{B}$, where $\bar{\mathcal{A}} \leq \bar{\mathcal{B}}$, where $\bar{\mathcal{A}}$ and $\bar{\mathcal{B}}$ denote the mean of elements in the clusters $\mathcal{A}$ and $\mathcal{B}$ respectively\;
		Cluster the elements of $\mathcal{A}$ into two clusters, $\mathcal{A}$ and $\mathcal{A}'$ such that $\bar{\mathcal{A}} < \bar{\mathcal{A}}'$\;
		\While{$|\bar{\mathcal{A}} - \bar{\mathcal{A}}'| > b$}{
			Cluster the elements of $\mathcal{A}$ into two clusters, $\mathcal{A}$ and $\mathcal{A}'$ such that $\bar{\mathcal{A}} < \bar{\mathcal{A}}'$\;
		}
		The number of elements remaining in $\mathcal{A}$ is the estimated number of true zero-valued elements, $n_{z}^{(s)}$, in $\boldsymbol{\theta}^{(s)}$
	}
	Find $\hat{n}_z =$ median value of $n_{z}$ \;
	Find $\hat{\boldsymbol{\theta}} =$ median values of the elements in $\boldsymbol{\theta}^{(1:S)}$  \;
	Set elements in $\hat{\boldsymbol{\theta}}$ with the $\hat{n}_z$ smallest absolute values to 0
	\caption{Sequential 2-means for posterior draws $s = 1,
		\ldots,S$ for parameter $\boldsymbol{\theta}$}
	\label{alg:sequential2means}
\end{algorithm}

\end{document}